\documentstyle[12pt,aasms4]{article}

\begin{document}
\oddsidemargin= 15mm 

\onecolumn

\title{The Type Ia Supernova 1997br in ESO576-G40}
{\center\author{W.D. Li\altaffilmark{1,2}, Y.L. Qiu\altaffilmark{1}, Q.Y. Qiao\altaffilmark{1}, X.H. Zhu\altaffilmark{1}, J.Y. Hu\altaffilmark{1},\noindent\newline M.W. Richmond\altaffilmark{3}, A.V. Filippenko\altaffilmark{2}, R.R. Treffers\altaffilmark{2}, C.Y. Peng\altaffilmark{4}, D.C. Leonard\altaffilmark{2}}}

\altaffiltext{1}{ Beijing Astronomical Observatory, Chinese Academy of Sciences,
Zhongguancun Beijing, 100080, P. R. China}

\altaffiltext{2}{Department of Astronomy, University of California, Berkeley, CA 94720-3411}

\altaffiltext{3}{Department of Physics, Rochester Institute of Technology, Rochester, NY 14623}

\altaffiltext{4}{Department of Astronomy, University of Arizona, AZ 85721}

\vspace{2cm}

\newpage

\begin{abstract}

The peculiar type Ia supernova SN 1997br in ESO576-G40 was extensively observed at
Beijing Astronomical Observatory (BAO) and Lick Observatory. In this paper, we 
present and discuss the $BVRI$ photometry and the spectra collected over
three months, starting 
from 9 days before maximum brightness. 

The light curves of SN 1997br are similar to those of SN 1991T, with slow decline
rates after the $B$ maximum. Well sampled data before the $B$ maximum show unambiguously that
SN 1997br rises more slowly and has a wider peak than normal type Ia supernovae. The optical 
color evolution of SN 1997br is also similar to that of SN 1991T. 
 We estimate the extinction of SN 1997br to be 
$E(B-V)$=0.35$\pm$0.10 by  comparing
its $BVRI$ light curves to those of SN 1991T and by measuring the equivalent width of 
interstellar Na I D absorption lines.

 We have conducted a thorough comparison
of the spectroscopic evolution of SN 1997br, SN 1991T, and SN 1994D. Although SN 1997br
is generally very similar to SN 1991T, it shows some interesting 
differences at various epochs. Spectra of SN 1997br seem to indicate an earlier transition to the dominant phase of Fe peak elements after the $B$ maximum. 
Si II lines in SN 1997br show a very short duration after the $B$ maximum.

We discuss the implications of our observations of SN 1997br for models of
type Ia supernovae. Specifically, we suggest that some SNe Ia may result from
decelerated detonations of white dwarfs.

\end{abstract}


\section{Introduction}

SN 1997br in ESO576-G40 was discovered by the BAO
Supernova (SN) Survey in an image taken on 1997 Apr 10.6 UT (Li et al. 1997a; Li 1997),
the twelfth supernova to be discovered by the BAO SN survey since its inception in early 1996.
A low resolution optical spectrum obtained shortly after the discovery (Qiao 
et al. 1997a) identified the event as a peculiar SN 1991T-like type Ia, with 
prominent Fe III absorption lines superimposed on a rather blue continuum.
Li et al. (1997a) measured an accurate position for SN 1997br as $\alpha$ = 13h20m42.s40,
$\delta$ = -22$^o$02'12.''3 (equinox J2000.0), which is 20.6'' west and 51.6'' north of the 
nucleus of ESO576-G40. A follow-up program of multicolor photometry and spectroscopy
was immediately established at BAO for this peculiar supernova. Numerous observations were also obtained at Lick Observatory. We report here on the
results from the first 90 days after discovery.

Section 2 contains a description of the observations and analysis of the photometry:
our methods of performing photometry, our calibration of the measurements onto
the standard Johnson-Cousins system, our resulting multicolor light curves, and our
comparison between the light curves of SN 1997br, SN 1991T, and other type Ia supernovae (SNe Ia).
Section 3 contains a description of the observations and analysis of the spectra
 of SN 1997br. The good  temporal coverage of the spectral 
observations (17 spectra from 9 days before to 81 days after $B$ maximum) enables us 
to monitor the evolution of many features. A through
comparison is done between the spectra of SN 1997br, SN 1991T, and the normal SN
 Ia 1994D. We discuss the implications of our observations in Section 4
and summarize our conclusions in Section 5.

\section{Photometry}

\subsection{Observations and data reduction}

Broadband $BVRI$ images of SN 1997br were obtained using CCD detectors on the 
BAO 0.6m telescope and the Lick 0.76m Katzman Automatic Imaging Telescope (KAIT) beginning the second day after discovery and 
continuing for about 90 days. 
The BAO observations were made at the f/4.23 prime focus using 
a TI 1024$\times$1024 pixel CCD (0.96 arcsec pixel$^{-1}$); only $BVR$ filters were used due to a shortage of 
positions on the filter wheel (the position of the $I$ filter was occupied by a clear filter 
for the BAO SN survey). The KAIT images ($BVRI$) were taken 
at the f/8.17 Cassegrain focus using a Tektronix 512$\times$512 pixel CCD (0.63 arcsec pixel$^{-1}$). 
Calibration of the field was done at BAO on 1997 Jun 14 by observing Landolt (1992)
standards at different airmass throughout the night, but reduction of the 
data suggests that conditions might not have been photometric.
Another calibration of the field came from observations made by Ricardo 
Covarrubias of CTIO, who used the CTIO 0.91m telescope on 1997 May 3, and we 
adopt it throughout this paper. Instrumental magnitudes for the standards were measured
using the IRAF DAOPHOT package and then used to determine
transformation coefficients to the standard systems of Johnson (Johnson et al. 1966; 
for $BV$) and Cousins (Cousins 1981; for $RI$). The derived transformation coefficients and color terms were used to calibrate a sequence of five stars near SN 1997br.
Figure 1 shows a KAIT image of SN 1997br, with the SN and the five local standard stars
marked. The magnitudes of these five stars and the associated uncertainties are listed
in Table 1. Star E is rather faint and did not always appear on all images.

Measuring the brightness of SNe is usually complicated by the background surface 
brightness of their host galaxies, necessitating template 
subtraction (e.g., Hamuy et al. 1994; Richmond et al. 1995). In 
the case of SN 1997br, however, the measurement is relatively straightforward:
the SN is 55.6 arcsec from the nucleus of its host galaxy, in a faint 
region of the host galaxy. 
We could have used simple aperture photometry to measure the magnitudes of the SN and 
the comparison stars, but we instead adopt the more sophisticated point spread function (PSF) fitting method for the following reasons:  
(1) Although the typical FWHM is about 3-4 pixels in the BAO
and KAIT images, some images have FWHM larger than 8 pixels, in which case
PSF fitting is superior. We choose to use the same method for all 
our images to ensure that the results are obtained in a consistent manner. 
(2) There
are many bright, isolated stars with which to define the PSF.  
(3) The PSF fitting method creates an image with
stars subtracted, from which we can estimate whether the brightness of the SN  
is measured correctly. 
 
Schmidt et al. (1993) found that the PSF fitting method can systematically overestimate
or underestimate the brightness of an object when there is a strong gradient in 
the background, but the errors can be reduced substantially by using only the inner
core of an object to fit the PSF. Though we don't have a strong gradient in the
background of SN 1997br, we choose to adopt this procedure  and set our
fitting radius of the PSF to be the value of FWHM. Using only the inner core to fit
the PSF also yields higher signal-to-noise ratios when performing differential photometry
between the SN and the local standard stars. 

To measure the sky background of the SN, we always use an annulus  
with radii 15-20 pixels (14.4''-19.2'' for BAO images, 9.45''-12.6'' for KAIT images), thereby avoiding possible substantial variations in
 the host galaxy light near the SN.   For local standard stars, on the other hand, we use an annulus with radius (4-5)$\times$FWHM (typically 14-20 pixels)
 to measure the sky background.

The instrumental magnitudes measured from this PSF fitting method are transformed onto 
the standard Johnson-Cousins system. For the BAO observations, the color terms of the filters 
are determined from calibration observations obtained in Dec 1995 and Jan 1996 and 
are of the following form:

$  V   = v - 0.014 (b-v) + C_V   $,

$ (B-V) = 1.19(b-v) + C_{BV}     $,

$ (V-R) = 0.91(v-r) + C_{VR}     $.
 
{\noindent Here, capital letters denote magnitudes in the standard system and lower-case
letters denote instrumental magnitudes. The terms $C_V,C_{BV}$ and $C_{VR}$ are the differences
between the zero points of the instrumental and standard magnitudes, which can vary from night
to night. Since we always measure the magnitudes of SN 1997br relative the those  of
several local comparison stars, these terms are not involved in the photometric
solutions.  The color terms for the KAIT observations are determined from 
calibration observations obtained on Jun 14, 15, and 18, 1998 and are of 
the following form:

$  V   = v - 0.005 (b-v) + C_V   $,

$ (B-V) = 1.12(b-v) + C_{BV}     $,

$ (V-R) = 0.83(v-r) + C_{VR}     $,

$ (V-I) = 0.95(v-i) + C_{VI}     $.

 }

Our final results of the CCD photometry are listed in Table 2, and our $BVRI$ light 
curves for SN 1997br are shown in Figure 2 together with the best $\chi^2$-minimizing fits to those of SN 1991T (Hamuy et al. 1996a for $BVI$;
 Ford et al. 1993 for R).
From these fits we can see that the photometric behavior of SN 1997br is very similar
to that of SN 1991T. After calculating the times and magnitudes
at peak, we will discuss each light curve in turn, comparing it with those of other type 
Ia SNe.

\subsection{Optical light curves}

The times and magnitudes of the peak in each filter are listed in Table 3. Unfortunately,
there are large gaps near the maxima of these light curves 
(especially $B$) because of poor weather and  the proximity
of the SN to the bright moon, so these parameters cannot be measured
directly. In order to determine these values we use two methods:
(1) fit a cubic spline function to the $BVR$ light curves (spline fitting), and 
(2) fit the $BVRI$ light curves of SN 1991T to those of SN 1997br (template fitting). Note that the second method is the only way to get an estimate of the time and 
peak of the $I$-band light curve since we have no pre-maximum observations.
We prefer to adopt the results of spline fitting for the $R$-band light curve, since 
the SN 1991T template is not a detailed match to the SN 1997br data and gives a peak 
0.08 mag brighter. 
Measurements for the $B$ and $V$ light curves from the two methods are 
consistent with each other within the uncertainties, so we average them to give the 
results in 
 Table 3. 

 In Figures 3-6, we compare the light curves of SN 1997br with those of several other well-observed SNe Ia: 
SN 1991T (Hamuy et al. 1996a for $BVI$; Ford
et al. 1993 for $R$); SN 1989B (Wells et al. 1994); SN 1994D (Richmond et al. 1995), and SN 1991bg
(Filippenko et al. 1992b; Leibundgut et al. 1993). All light curves are shifted in times and peak magnitudes to match those of
SN 1997br.

The $B$-band light curve of SN 1997br (Fig. 3) is similar to, but seems to decline  faster than that of SN 1991T. Following Phillips (1993) and 
Hamuy et al. (1996a), we define $\Delta m_{15}(X)$ as the difference between the magnitude at peak and 15 days 
after the peak in the $X$ passband, and measure  $\Delta m_{15}(B)$ = 1.00$\pm$0.15 for SN 1997br.
The uncertainty, which is somewhat
large due to our lack of observations around the peak, is the sum in quadrature of the uncertainty in the $B$-band
peak magnitude, the change in the interpolated magnitude 15 days after peak if the date 
of maximum is changed by the amount listed in Table 3, and the uncertainty of the observed 
magnitudes. This $\Delta m_{15}(B)$ value of SN 1997br is larger than that of SN 1991T (0.94$\pm$0.07),
but smaller that that of SN 1989B (1.31$\pm$0.07), SN 1994D (1.28$\pm$0.05),
and SN 1991bg (1.88$\pm$0.10), and is at the lower end of the distribution of $\Delta m_{15}(B)$
for all SNe Ia in the Cal\'an/Tololo Supernova Survey (Hamuy et al. 1996b). The decline rate of SN 1997br changes at about 30 days past maximum and is nearly the 
same as that of SN 1991T thereafter.

The behavior of such a 1991T-like type Ia SN is obtained 
 with good quality CCD measurements at a very early phase (9 days before
$B$ maximum). These premaximum observations of SN 1997br
show unambiguously that it rises more slowly than normal SNe Ia such as SN 1994D and may
have a larger rise time from explosion to $B$ maximum. The width of the peak at 0.6 mag
below maximum is about 20 days for SN 1997br, while it is about 17 days for SN 1994D.

In the $V$ band (Fig. 4), the peak width of SN 1997br is again similar to that of SN 1991T,
and is wider than that of SN 1994D, SN 1989B, and SN 1991bg. The $V$ light curve of SN 1997br
displays a wide, flat-topped feature around it maximum, within 0.05 of the peak for 
more than 10 days. Using the definition of $\Delta m_{15}$ above, we measured $\Delta m_{15}(V)$=0.57$\pm$0.08 for SN 1997br, which
is similar to that of SN 1991T (0.59$\pm$0.06) and smaller than that of SN 1994D (0.76$\pm$0.04) and SN
1989B (0.71$\pm$0.07).

We show the $R$ band light curve in Fig. 5, in which SN 1997br begins to deviate strongly from
a simple decline after $B$ maximum. From 13 to 44 days after $B$ maximum,the $R$-band light curve 
develops a prominent ``shoulder'' seen in most other well-observed SNe Ia in the $R$-band (Ford et al. 1993; Wells et al. 1994) except SN 1991bg. After this the curve declines at a constant rate of about 
0.04 mag per day. SN 1997br shows a wider peak in the $R$ band than do SN 1994D, SN 1989B, 
and SN 1991bg, and  a longer duration ``shoulder" than do SN 1994D and SN 1989B.  The fit to the $R$-band light curve of SN 1991T is not so good,
in part because the light curve of SN 1991T is not well-sampled.  
In particular, 
the lack of a ``shoulder" in the $R$ light curve of SN 1991T is due to a
paucity  of observations
at the period.   

Our $I$-band light curve of SN 1997br is shown in Fig. 6. Unfortunately there     aren't $I$-band observations
at BAO, and the $I$-band observations from KAIT only began 6 days after the $B$ maximum, so we are unable
to directly measure  the time and magnitude of the $I$-band peak. An estimate is done by fitting the $I$-band
light curve of SN 1991T to that of SN 1997br; this yields the values listed in Table 3. From the fit we can
see that the photometric behavior of SN 1997br in the $I$-band is almost the same as that of SN 1991T
over the period of overlap. Unlike SN 1994D, which displays a pronounced dip 
at about 14 days after $B$ maximum and a secondary peak about 23 days after $B$ maximum, SN 1997br
and SN 1991T show less obvious minima and secondary maxima; instead, there is a 
nearly constant plateau in $I$-band magnitude from about 14 to 24 days after $B$ maximum. 

In summary, the light curves of SN 1997br closely resemble those of SN 1991T, and differ significantly
from those of normal SNe Ia such as SN 1994D and SN 1989B. The well-sampled premaximum light curves
of SN 1997br unambiguously show that it rises  and declines more slowly than 
do normal SNe Ia, and thus has wider peaks in the $BVR$ passbands.   

\subsection{Optical color curves}

We present the optical color curves $B-V, V-R$ and $V-I$, for SN 1997br in Figs. 7-9. We also show data from several other SNe Ia (1991T, 1989B, 1994D, and 1991bg) for
comparison. We correct the SN 1997br data for a color excess of $E(B-V)$=0.35 mag (which we derive 
below), and follow Wells et al. (1994) in correcting the data for SN 1989B with 
 $E(B-V)$=0.37 mag. A reddening of $E(B-V)$=0.04 is assumed for SN 1994D (Richmond et al. 1995). 
There is no reddening for SN 1991bg (Filippenko et al. 1992b).
To calculate the extinction in the $B,V,R$, and $I$ passbands we adopt the average interstellar extinction 
curve given in Table 2 of the review by Savage \& Mathis (1979). In the discussion below,
we denote any dereddened quantity with a superscript zero, such as $(B-V)^0$; values which lack 
the superscript are not corrected for reddening.

The $B-V$ color of SN 1997br shows a well-defined curve that conforms fairly well to those of SNe 1991T,
1989B, and 1994D, but differs significantly from that of the peculiar,
 subluminous SN 1991bg. At the time of the
discovery (-9 days), SN 1997br has an observed color $B-V$=0.31 mag,corresponding to an intrinsic
color $(B-V)^0$ = -0.04 mag for $E(B-V)$=0.35 mag. It then becomes progressively bluer, reaching
a minimum value of $(B-V)^0=-0.14$ mag at 4 days before the $B$ maximum, and reddens again after this. 
At the time of $B$ maximum the color is estimated to be $(B-V)^0$=-0.03.   
 After $B$ maximum the $B-V$ color of SN 1997br 
becomes progressively redder than that of SN 1991T ($\Delta(B-V)\approx$0.12 mag
at 25 days),  because SN 1997br has a more rapid decline rate in the $B$-band after its maximum.  
Both SN 1991T and SN 1997br reach about the same $B-V$ color, $(B-V)^0 \approx$  1.1, about 30 days after 
maximum. SN 1994D and 1989B seem to reach their peaks earlier (25 days after $B$  maximum). Subsequently, the
$B-V$ colors for all SNe Ia (including SN 1991bg) turn slowly to the blue again.

The $V-R$ color of SN 1997br in Fig. 8 shows a similar pattern of evolution as those of 
SNe 1989B, 1991T and 1994D, but seems slightly bluer before 35 days after maximum. The agreement between the $V-R$ color of SN 1997br and that of SN 
1991T is quite good, except during the time from -10 days before and 
10 days after B maximum, in which time SN 1991T had very sparse observations.

The $V-I$ colors of these five SNe Ia (Fig. 9) show more variations than do  the $B-V$ and $V-R$ colors.
It is thus striking that the curve of SN 1997br is so similar to that of SN 1991T.
Different SNe seem to reach their reddest $V-I$  colors at different epochs: SN 1991bg
does it at day 17, SN 1994D at day 25, SN 1989B at day 35, and SNe 1997br and 1991T at day
40.

In general, all $B-V$,$V-R$, and $V-I$ color curves have similar a blue--red--blue pattern of
evolution. Each color curve have a minimum and a maximum, but the times of these minima
and maxima differ significantly. The $B-V$ color reaches its bluest value at about day
-4 and its reddest value at about day 25-30; the $V-R$ and $V-I$ colors do this 
at day 10 and day 20-40, respectively.

\subsection{Extinction}

Conspicuous interstellar  Na I D absorption lines are present in our low-resolution spectra of SN 1997br (Sec. 3), implying substantial reddening by intervening 
material in the Milky Way and in the host galaxy ESO576-G40.
In this section we will attempt to derive the amount of interstellar extinction 
towards SN 1997br in two ways: by comparing its optical colors with those of 
other type Ia SNe, and by measuring the strength of interstellar absorption 
features in its spectra. 

Considering the similar photometric behavior of SN 1997br and SN 1991T, we 
might assume that they should have similar color evolution and intrinsic 
colors at maximum light. We can then measure the relative
amount of extinction between them very easily. Comparing the colors at the time
of maximum in the $B$-band, we find differences of 

($B-V$)$_{97br}$ - ($B-V$)$_{91T}$ = 0.19,

($V-R$)$_{97br}$ - ($V-R$)$_{91T}$ = 0.14, and

($V-I$)$_{97br}$ - ($V-I$)$_{91T}$ = 0.35 mag.

{\noindent With our adopted extinction coefficients these values imply that  $\Delta E(B-V)$
equals 0.19, 0.21, 0.22 mag, respectively, between SN 1997br and 1991T.}

The reddening of SN 1991T, however, is uncertain. Filippenko et
al. (1992a) and Ruiz-Lapuente 
et al. (1992) estimated $E(B-V)$ values of 0.13 and 0.34 mag, respectively,
 based on interstellar Na I D lines in the parent galaxy.
Phillips et al. (1993) estimated a reddening of $E(B-V)$=0.13 for SN 1991T based on the assumption that 
type Ia SNe share the same intrinsic $(B-V)$ color of 0.0 mag at $B$ maximum.
Jeffery et al. (1992), on the other hand, ruled out a reddening larger than 0.2 mag
based on comparisons of spectra of SN 1991T and 1990N.
Adopting $E(B-V)$ = 0.13 mag for SN 1991T, 
we get $E(B-V)$ values of 0.32, 0.34,  and 0.35 mag for SN 1997br from our color excesses
in $B-V$, $V-R$, and $V-I$ respectively. 

In our low-resolution spectra for SN 1997br we measured the total equivalent width (EW) 
of the Na I D lines (including the Galactic component) to be EW = 1.85 $\pm$ 0.1 \AA .
Adopting the prescription discussed by Filippenko et al. (1990) for SN 1987M, 
we derive a reddening of $E(B-V)$= 0.39 mag for SN 1997br. This estimate, however, is quite 
uncertain. For Galactic stars, the uncertainty in the relationship between 
the Na I D EW and $E(B-V)$ is large (e.g., Hobbs 1978). Although the conversion
factor adopted by Filippenko et al. (1990) generally appears to underestimate the 
extinction, Jeffery et al. (1991) pointed out that this approach might 
actually overestimate the extinction. In any case, we consider the EW of the 
interstellar Na I D lines to be only an approximate indicator of the reddening.

In this paper we adopt a reddening of $E(B-V)$ = 0.35 $\pm$0.10 for SN 1997br.

\section{Spectroscopy}

\subsection{Observations and Reductions}

The spectroscopic observations were done mostly with the BAO 2.16-m telescope 
using the Photometrics spectrograph equipped with a Tektronix 1024 x 1024
pixel CCD. We also report here the spectra of SN 1997br obtained 
with the Lick 3-m Shane telescope (+ Kast  CCD spectrograph; Miller \& Stone 1993). The journal of observations is given in Table 4.

All of the optical spectra were reduced and calibrated using standard techniques. Wavelength calibrations were carried out via comparison Fe-Ar or He 
lamp exposures taken at the position of the supernova. The spectra were corrected for continuum  
atmospheric extinction using mean extinction curves for BAO and Lick;
also, telluric lines were removed from the Lick (but not BAO) data. The final flux calibrations were derived from observations of one or more spectrophotometric
standard stars (Oke \& Gunn 1983). The spectra were obtained at large airmasses because of  the object's southern declination (-22$^o$). For spectra obtained
at Lick, particular care was  taken to always align the slit along the parallactic angle, to avoid chromatic effects in the data (Filippenko 1982). 
This procedure was not adopted at BAO, however, and it may cause some problems 
for the continuum shape of the spectra. Comparison of 
data obtained at BAO and Lick at similar epochs (e.g., the two spectra at day -4
 in Fig. 10), however, reveals significant differences only at the extreme blue end, so this problem is not serious.  

Our final set of spectra covering the period from 9 days before to 81 days after
$B$ maximum is displayed in Fig. 10. The first spectrum we obtained at BAO 9 days
before maximum (Apr 11 UT) clearly illustrate the pecularities initially reported 
for SN 1991T -- the blue and nearly featureless continuum, and the two major  absorption lines at approximately 4270 and 4975 \AA.  The biggest anomaly, however, is the absence of 
Si II $\lambda$6355 and the Ca II H \& K lines. The latter features are barely 
visible in the spectrum obtained on Apr 
16 UT (4 days before $B$ maximum). 

As with the photometry, there is a big gap from 4 days before to 8 days 
after $B$ maximum because of poor weather and
the object's proximity to the bright moon. During this period dramatic changes 
occur, so the spectrum at 
day +8 is quite different from that at day -4. The continuum becomes 
considerably redder, and conspicuous lines of Si II $\lambda$6355, Ca II 
(H, K and IR triplet),
and other intermediate-mass elements (S, Mg, etc.) appear in
the day +8 spectrum. The spectrum now clearly 
resembles that of a Type Ia supernova, although the strengths of all
intermediate-mass element absorption lines are relatively weaker than 
in normal SNe Ia. These data leave no doubt that SN 1997br is a genuine (though peculiar) type Ia 
event like SN 1991T. 

Of particular interest is the evolution of Si II $\lambda$6355,
which is the defining feature of type Ia SNe.  
Spectra taken earlier than 4 days before $B$ 
maximum don't show 
significant evidence of this line (Fig. 10). A weak Si II $\lambda$6355 line
appears in the day -4 spectrum, showing a broad absorption (FWHM 9000 km/s)
centered at 6150 \AA\, and with the blue edge corresponding to a velocity of
 18000 km/s. Although this value is smaller than that measured in SN 1990N
on day -14 (25000 km/s; Leibundgut et al. 1991) and in SN 1994D on day -11 (21600 km/s;
Patat et al. 1996), it still indicates that the Si synthesized during the 
explosion extends well beyond the limit fixed by the unmixed W7 deflagration
model (15000 km/s; Nomoto et al. 1984), as pointed out by several other
authors (Mazzali et al. 1993; Kirshner et al. 1993). 

In the second spectrum (day +8), the Si II $\lambda$6355 absorption 
line appears well developed, but considerably
narrower (FWHM 6500 km/s), with a blue edge corresponding to a 
velocity of 15500 km/s. The absorption component moves slightly to the red 
in the spectra of days +10 and +12. Meanwhile, other features appear and contaminate the Si II line, most notably its red side (probably Fe II 
$\lambda\lambda$6456, 6518 at $\sim$ 6270 \AA). Blueward of the line, 
there may be some Fe II $\lambda\lambda$6238, 6248 at $\sim$ 6050 \AA. 
These changes can be seen most dramatically in the day +21 and +24 spectra, 
where the Si II $\lambda$6355 line is heavily obscured by these Fe II lines, 
and it disappears thereafter. 

This behavior of the Si II $\lambda$6355 line doesn't follow that defined by the template 
spectral evolution of typical SNe Ia like SN 1989B (Wells et al. 1994)
and SN 1994D (Patat et al. 1995; Filippenko 1997), which show a strong Si II $\lambda$6355 
line at times well before and near $B$ maximum.  The later appearance, the earlier disappearance, 
and the general weakness of 
the Si II $\lambda$6355 line in SN 1997br suggest that the Si layers in 
the ejecta of the SN  may be confined to a 
very narrow velocity region, and the abundance of Si in the ejecta may 
be less  than normal, as we will further discuss below.

To do a thorough comparison between the spectroscopic behavior of SN 1997br, SN 1991T,  
and typical 
SNe Ia like SN 1994D, we plot the spectra of these 3 SNe at different
epochs (days -9, -3,+10,+14,+21,+51) in Figs. 11 to 16.
All the spectra  have been dereddened ($E(B-V)$ = 0.35 mag for SN 1997br;
$E(B-V)$ = 0.13 mag for SN 1991T;  $E(B-V)$ = 0.04 mag for SN 1994D as derived by Richmond
et al. 1995) and have been corrected for the observed galaxy redshifts.
The line identifications adopted here are taken from Kirshner et al. 
(1993), Jeffery et al. (1992), and Mazzali et al. (1995). 

At times before and around maximum light the SN Ia spectrum is produced
by the outermost material in the ejecta and is composed
mainly of lines of intermediate-mass elements for normal cases (Branch et al. 1985; Kirshner
et al. 1993). This can be seen clearly in the day -9 spectrum of 
SN 1994D in Fig. 11; it is dominated 
by lines of Si II, S II, Mg II, Ca II, and O I. The spectra of
SN 1997br and SN 1991T are similar to each other but significantly
different from that of SN 1994D. The two main absorption features are believed to be 
multiplets of Fe III $\lambda$4404 and Fe III $\lambda$5129, respectively (Jeffery 
et al. 1992; Ruiz-Lapuente et al. 1992). This peculiar phenomenon
of Fe-dominated type Ia pre-maximum spectra is explained by 
many authours in terms of abnormal abundances and/or 
abnormal temperatures in the ejecta of the SN (Filippenko 
et al. 1992a; Ruiz-Lapuente et al. 1992; Phillips et al. 1992; 
Jeffery et al. 1992).

By day -3, the spectrum of SN 1994D has changed significantly. Si II $\lambda$3858 is well
separated from the Ca II H \& K lines, and is stronger, as is the Si II $\lambda$4560 line. Si II $\lambda$6355 now 
has a well-developed P-Cygni profile. S II $\lambda$5468 and S II 
$\lambda\lambda$5612, 5654 form the characteristic ``W"-shaped feature 
in the 5200 -- 5600 \AA\, region. The Ca II H \& K lines, however,
are less prominent than in the day -9 spectrum. 
The spectra of SN 1997br and SN 1991T are still dramatically different
from that of SN 1994D. In the day -4 spectrum of SN 1997br,
the characteristic feature of SNe Ia, Si II $\lambda$6355, appears only 
marginally at 6100 \AA, accompanied by other Si lines at 4400 \AA\, (Si III $\lambda$4560), 4800 \AA\, (Si II $\lambda$5051), and probably 
 3700 \AA\, (Si II $\lambda$3858). 
There is also a hint of the Ca II H \& K lines at 3800 \AA. S II 
  (the ``W" shape) may be responsible for the weak flat-bottomed
absorption in the 5200 -- 5400 \AA\, region. Our day -3  spectrum of SN 1991T 
doesn't extend out to the Si II $\lambda$6355 line, but Phillips et al. (1992) report
that Si II $\lambda$6355 line was weakly present at this epoch, and in most
respects the spectrum is quite similar to that of SN 1997br. 

The spectra of these 3 SNe about 10 days after $B$ maximum (Fig. 13) differ from 
the premaximum ones. As marked in Fig. 13, Fe II and Co II
lines begin to appear in the spectrum of SN 1994D, and Na I D contributes to the 
absorption at 5750 \AA. O I $\lambda$7773 and the Ca II IR triplet 
become much stronger, while the ``W"-shaped S II lines
become much weaker than in the day -3 spectrum. More significant changes 
happen to the spectra of SN 1997br and SN 1991T: the appearance of their 
spectra is now generally similar to that of SN 1994D. Intermediate-mass 
element lines of Si, O, Ca, and S appear in their spectra, although their intensities
are relatively  weaker than in SN 1994D. Fe-peak element lines,
on the other hand, are better developed in the spectra of SN 1997br and SN 
1991T. Comparison of the spectra of SN 1997br and SN 1991T at this epoch
reveals minor yet interesting differences -- the Si II and the ``W" shaped S II lines are slightly stronger
in SN 1991T, while the Fe II and Co II lines are better developed in
 SN 1997br. The red wing of the Si II $\lambda$6355 line of SN 1997br,
even at this phase, shows contamination by Fe II $\lambda\lambda$6456, 6518. This 
might indicate that spectra of SN 1997br have a shorter duration of intermediate-mass element
lines and a faster transition to the Fe-peak element lines 
after $B$ maximum than do those of SN 1991T.

More evidence for this  
can be found in the day +14 spectra in Fig. 14, where the Si II $\lambda$6355 line
of SN 1997br shows a flat-bottomed absorption due to the increased
strength of Fe II, while those of SN 1991T and SN 1994D still retain a well-defined P-Cygni profile. As marked in Fig. 14, the spectrum of SN 1997br also shows stronger evidence 
of Fe II $\lambda\lambda$6238, 6248 and Co II $\lambda$4152 absorption.
The three spectra exhibit quite different features around the 5200 -- 5600 \AA\, region.
SN 1994D shows a notch of absorption at 5340 \AA\, that might be caused
by Fe II $\lambda$5535; SN 1997br also shows 
this absorption and possibly its emission component at around 5500 \AA. 
SN 1991T, however, doesn't show this absorption, and may still have a hint 
of the ``W"-shaped S II lines.

Inspection of the day +21 spectra of the 
three SNe (Fig. 15) reveals that this trend continues. Fe II and Co II lines
now dominate all three spectra. SN 1994D still shows a distinct Si II $\lambda$6355 line, while SN 1991T may have a hint of its existence in the flat-bottomed
absorption at around 6100 \AA. SN 1997br, on the other hand, is dominated
by Fe II $\lambda\lambda$6456, 6518 lines. In the 5200--5600 \AA\, region,
both SN 1994D and SN 1997br exhibit absorption at 5300 \AA\, and prominent emission
at 5500 \AA, while SN 1991T only has the absorption. All three spectra show 
quite similar features in the region blueward of 5400 \AA, although it can be seen
that the emission lines at 4600 \AA\, (Fe II $\lambda$4555) and 5000 
\AA\, (Fe II $\lambda$5018) have comparable strengths in SN 1994D, while 
the 4600 \AA\, emission  is stronger than that at 5000 \AA\, in SN 1997br and in SN 1991T.

Given all the differences we have seen in the spectral evolution of these 
three SNe so far, the overall similarity of their nebular phase spectra (about 50 days after B maximum 
in Fig. 16) is quite impressive. The difference in
the 5200 -- 5600 \AA\, region disappears: all of them show strong emission
at 5500 \AA. The intensity ratios of the emission lines at 4600 \AA\, (Fe II $\lambda$4555) to that at 5000 \AA\, (Fe II $\lambda$5018)  are similar. 
Careful inspection of the spectra reveals, however, some minor 
difference among them. Even at this late phase, SN 1994D still shows a weak
Si II $\lambda$6355 line at 6100 \AA. The Na I D line is also stronger in
SN 1994D than it is in SN 1997br and SN 1991T. More evidence of differences
between the spectra can be seen at around 4800 \AA, 5400 \AA, 6000 \AA, 
and 7000 \AA.

Our comparison shows clearly that the peculiarity of the spectroscopic behavior of SN 1997br and SN 1991T is
not confined only to before and around maximum as reported by Phillips et al. (1992),
but extends to the nebular phase. At any given epoch, SN 1994D (normal SN Ia) always shows stronger intermediate-mass element lines, while SN 1997br and SN 
1991T show stronger and better developed Fe-peak element lines. 

Moreover, although the overall spectroscopic behavior of SN
1997br is similar to that of SN 1991T, there are still important differences.
Specifically, SN 1997br has a faster 
transition to Fe-peak element lines in the nebular phase. If we explain the peculiar
spectroscopic behavior 
of these two SNe in terms of their abundance anomaly,  this might mean that 
SN 1997br has shallower and thinner layers of intermediate-mass  elements  in the ejecta than does SN 1991T. 

An abundance anomaly, however, might not be the only way to account for the 
peculiarities and differences seen in the spectra of SN 1997br and SN 1991T:  
different temperature evolution of the photosphere  may play some
role. As can be seen in the pre-maximum spectra (Figs. 11 and 12), SN 1997br and SN 1991T are somewhat bluer than SN 1994D and thus may have higher photosphere 
temperatures. This higher temperature favors the production of Fe III lines but not of intermediate-mass element lines (Mazzali et al. 1995).
As the SNe evolve, their photospheric temperatures drop, and intermediate-mass
element lines begin to form.  Evidence that SN 1997br had a faster 
drop in temperature than SN 1991T can be found in Fig. 13, 14, and 15, where the continuum of SN 1997br is redder than that of SN 1991T and is similar to that of SN 1994D.
The slower temperature evolution of SN 1991T
may also be responsible for its different evolution in 5200--5600 \AA\, region
on days +14 and +20.
The bluer continuum of SN 1991T compared with that of SN 1997br on  days +14 
and +20 
is also consistent with the photometric behavior of these two SNe seen in 
Fig. 7:  SN 1991T is bluer at epoch 10 to 30 days after $B$ maximum in the 
($B-V$) color. 

The expansion velocities as inferred from observed minima in the spectra 
may provide some clue to the nature of SN Ia explosions (Branch et al. 1988). 
Fig. 17 shows the derived expansion velocities of the Fe III $\lambda$4404,
Fe III $\lambda$5129, Fe II $\lambda$4555, and Si II $\lambda$6355 lines
for SN 1997br and SN 1991T, together with those derived 
from Si II $\lambda$6355 for SN 1981B (Branch et al. 1983), SN 1986G
(Phillips et al. 1987), SN 1989B (Wells et al. 1994), SN 1991bg (Leibundgut
et al. 1993), and SN 1994D (Patat et al. 1996). Comparison of data reveals that the expansion velocities
of the Fe lines (Fe III $\lambda$4404, Fe III $\lambda$5129, Fe II $\lambda$4555) are slightly higher in SN 1991T than in SN 1997br, while the velocities
of Si II $\lambda$6355  are higher in SN 1997br than in SN 1991T.
This indicates that SN 1997br indeed has a relatively shallower (and thus 
higher velocity) layer of
Si compared to SN 1991T, consistent with our earlier discussion.
The Si II velocities for SN 1997br are remarkable in  their slow
decline (-11500 km/s to -10500 km/s)  over the 22 day period (-4 to +18
days) during which this line was measurable.
SN 1991T shows similar behavior, as reported by Phillips et al. (1992),
 and this might imply that the Si zone in the
expanding ejecta of SN 1997br was confined to a fairly restricted 
layer characterized by velocities in the range 10500--11500 km/s 
(for SN 1991T the velocity range is 9000--10000 km/s).

The  Fe III $\lambda$4404 and Fe III $\lambda$5129
line velocities show an abrupt change in the decline rate around 
day -7. A similar phenomenon has been found in SN 1990N (Leibundgut
et al. 1991) and in SN 1994D (Patat et al. 1995) for the Si II $\lambda$6355 line, and was explained in terms of changes in the density gradient
and increased amounts of absorbing layers as the photosphere recedes.

\section{Discussion}   

\subsection{Absolute magnitude of SN 1997br}

Having measured the peak magnitude of each light curve, and found a
value for the extinction, we lack only one piece of information needed
to calculate the absolute magnitude of SN 1997br: the distance to 
ESO576-G40. 

This galaxy, also named MCG-04-32-007 and PGC 46574,
is listed as a member of one group of 
galaxies, namely group 345 of the Lyon Group of Galaxies (LGG) Catalog
(Garcia 1993). Other members  and their heliocentric recession 
velocities  are NGC 5084 (1514 km/s), NGC 5087 (1625 km/s), NGC 5134 (1550 km/s), and ESO 576-G50 (1781 km/s). The heliocentric 
velocity of ESO576-G40 measured by several groups using
different methods (e.g., Fairall et al. 1992; Han 1992; Mathewson 
et al. 1992; Fouque et al. 1990)  yields values between 2055 and 2086
km/s. The luminosity-weighted mean recession velocity of the 
group is calculated to be 1583 km/s (Garcia 1993).

Zeilinger, Galleta, \& Madsen (1990)  
found that ESO576-G40 appeared to be perturbed by the presence
of NGC5084 in their optical image. Carignan et al. (1997) further
suggested that NGC 5084, as one of the most massive disk galaxies known,
had survived the accretion of several satellite galaxies similar to ESO576-G40. All these studies suggest that ESO576-G40 
may have a physical relationship with NGC 5084 and thus be at the
same distance. In this sense we can use the recession velocity 
for the group, 1583 km/s, as a distance measurement for 
ESO576-G40.

Mathewson et al. (1992) presented photometric and spectroscopic
observations of 1355 southern spiral galaxies and determined their
distances and peculiar velocities via the Tully-Fisher (TF) relation.
All their distances are measured relative to the Fornax cluster 
whose distance was taken to be 1340 km/s. They measured 
recession velocities of 2085 km/s (heliocentric) and 2403 km/s 
(CMB frame)  for ESO576-G40. 
The TF distance to ESO576-G40 was measured to be 1248 km/s after the 
Malmquist bias correction. This distance, with a peculiar velocity
of 1155 km/s relative to the CMB frame, may have large associated uncertainties
because ESO576-G40 is a highly inclined (nearly edge-on) galaxy
and is not well measured with the TF relation, so we are reluctant 
to use this distance measurement for ESO576-G40.

Adopting a distance of 1583 km/s for ESO576-G40 and a Hubble constant of
65 km s$^{-1}$ Mpc$^{-1}$ (Hamuy et al. 1996), the extinction-corrected 
absolute magnitudes for SN 1997br are listed below,

      $M_B$ = -19.29$\pm$0.42,

      $M_V$ = -19.28$\pm$0.32,

    $M_R$ = -19.02$\pm$0.25,

     $M_I$ = -18.97$\pm$0.18.

{\noindent The errors of the magnitudes are the sums in quadrature of the uncertainties
in peak magnitudes and the uncertainties in extinctions.}

From the absolution magnitudes we can see that  SN 1997br may be somewhat
more luminous than typical SNe Ia. This is consistent with the
fact that SN 1997br shares many similarities with the overluminous
SN 1991T. The absence of  a better distance measurement for its host galaxy, however, 
prevents us from making  definitive conclusions regarding  the absolute magnitudes
of SN 1997br.

\subsection{Heterogeneity of type Ia SNe}

It is well known that SNe Ia display a remarkable degree of similarity, 
more than any other subclass of SNe, and their spectral evolution and light
curves follow a reproducible pattern (Minkowski 1939; Kirshner et al. 1973;
Wells et al. 1994). During the past 15 years, however, spectroscopic and photometric
peculiarities have been reported with increasing frequency in well-observed SNe Ia, 
 and it is now settled that  SNe  Ia are not perfectly homogeneous
( see Filippenko 1997 for a review).
 
Nevertheless, Branch et al. (1993) conclude that the observational sample of SNe Ia is 
strongly peaked at spectroscopically normal (89\% in a sample of 84 SNe
Ia studied). We emphasize here, however, that this rate may be overestimated since
most of the spectra of these SNe Ia were taken at or after their optical
maxima, while the peculiarity of some SNe Ia  is most conspicuous before their optical
maxima (e.g., SN 1991T and SN 1997br). They also did not account for the observational
bias against subluminous SNe Ia like SN 1991bg. 

A striking statistic 
from the BAO SN survey
is that, among the 12 SNe Ia discovered in 1.5 years,
4 (33\%) are peculiar. SN 1997br (Li et al 1997a), SN 1997cw (Qiao et al. 
1997b), and SN 1998ab (Wei 1998) are SN 1991T-like, while SN 1997cn (Li et al.
1997b) is SN 1991bg-like. Nearly all SNe Ia discovered by the BAO SN survey
were found before their optical maximum. Though the SN Ia sample from
the BAO SN survey is small and our derived rate (33\%) is subject to
small-number statistics, it does suggest that the frequency of peculiar SNe Ia 
might be larger than that reported by Branch et al. (1993), and it stresses the importance
of discovering and observing these SNe before their optical maxima.

\subsection{Models of SNe Ia}

It is generally accepted that SNe Ia are thermonuclear explosions of carbon-oxygen 
white dwarfs with mass very near the Chandrasekhar mass ($M_{ch}$) of 1.4 $M_\odot$.
From the theoretical standpoint the key question is how the 
burning front or flame propagates through the white dwarf. Models of 
SNe Ia proposed in the past three decades can be divided into
three classes: deflagration (turbulent convective subsonic propagation;
e.g., Nomoto et al. 1976), detonation (a supersonic shock wave that initiates burning; 
e.g., Arnett 1969), and delayed detonation (the flame starts
as a deflagration and accelerates into a detonation later; e.g., Khokhlov 1991).

Recently a large variety of models of these three classes have been investigated by several  
papers of
H\"oflich, Khokhlov, and Muller (see references in H\"oflich et al. 1995). They
conclude that no single model is capable of representing
the whole range of SNe Ia. 
In general, pure detonation models burn the entire star
into iron-peak elements and can be ruled out for SNe Ia because of the presence of 
significant intermediate-mass elements. Diffferent variations of the delayed 
detonation models, on the other hand, can be used to explain various SN Ia 
light curves (but not necessarily the explosions themselves). 
They are also used to explain 
the peculiar behavior of Fe-dominated pre-maximum spectra of SN 1991T 
(Filippenko et al. 1992a; Ruiz-Lapuente et al. 1992). Deflagration 
models may be necessary to explain subluminous SNe Ia like SN 1991bg (Filippenko et al. 1992b; Leibundgut et al. 1993). 

Despite the successes of these models, the physics of the explosion is far from being
completely solved.
All of them need to be parameterized to avoid intractable 
three-dimensional calculations, and  none  has a rise time long enough to be consistent with the observations (e.g., $>$ 17.5 days for SN 1990N; Leibundgut et al. 1991). The initial explosive state of the 
white dwarf and the history leading to this state are not well 
understood (see reviews by, e.g., Woosley \& Weaver 1986; Wheeler \& Harkness
1990; Branch et al. 1995).

The physical conditions of different  white dwarfs at the time of explosion may be 
dissimilar, and this could cause the heterogeneity seen in SNe Ia. Possible differences
in the physics of the white dwarfs are as follows,

1) $Composition$. A helium  white dwarf will have a different explosion than a 
carbon-oxygen white dwarf. 

2) $Temperature$.  Variations in temperature may be caused by different accretion rates and different compositions
of the accreted matter from the companion star.

3) $Magnetic\,field$. No models that account for the magnetic fields of white dwarfs are 
available, yet they might affect the explosion considerably. 

4) $Rotation$. No models that account for the rotation of white dwarfs are 
available either, yet rotation is known to change the explosion characteristics of Type II SNe
considerably (e.g., M\"{o}nchmeyer 1990).

5) $Mass$.  Because of differences in  composition, temperature, magnetic field,
  rotation, and other factors of the white dwarf,  the Chandrasekhar mass 
is not a unique number, and  
 explosions may occur at different masses. At least one SN Ia, SN 1991bg,  may have exploded at a much smaller mass than 1.4 $M_\odot$ (Filippenko et al. 1992b; Leibundgut et al. 1993).

The outcome of a $decelerated\,detonation$ (i.e., a detonation initializes the explosion of 
a white dwarf, then decelerates and turns into a deflagration) has not been
carefully studied by theorists, yet it might be a promising model for some
SNe Ia (normal or SN 1991T-like). When the detonation propagates outward, it 
may decelerate because of decreasing matter density (less burning
material density), decreasing temperature (more energy needs to be
absorbed to ignite burning), and whatever other reasons. We propose this
model for the following reasons:

(1) It may provide a model for normal SNe Ia. The outer part of the ejecta will consist of a deflagration composition
that has been favored by spectral simulations. Moreover,  
the expansion velocity will be higher than  that of  pure deflagration models like W7
because of the detonation origin. This is consistent with observations of 
several recent SNe Ia, which require higher outermost expansion velocities
than model W7 (e.g., Leibundgut et al. 1991; Patat et al. 1996; Kirshner et al. 1993).

(2) It may also account for SN 1991T-like SNe Ia. When the initial 
detonation is energetic enough, it will propagate as a detonation until 
reaching the outermost part of the envelope, where it can become a 
deflagration. Events like
this may have a higher luminosity, higher photospheric temperature, and
less mass consisting of intermediate-mass elements in the outermost envelope, and may 
result in events like SN 1991T and SN 1997br.

The deceleration in our proposed  model is
in direct contrast to the acceleration inherent in  the very 
successful delayed detonation model,
 which has been widely accepted and used. 
However, there isn't rigorous physical support for the mechanism of acceleration 
or deceleration, so the outcomes of both scenarios should be investigated;
 the decelerated detonation model may 
yield some fruitful results.

\section{Conclusions}

In this paper we reported our photometric and spectroscopic observations of the peculiar 
type Ia SN 1997br in ESO576-G40. The overall photometric 
and spectroscopic behavior of SN 1997br closely resembles that of the peculiar type 
Ia SN 1991T, but with some noticeable differences.
 
The $BVRI$ light curves of SN 1997br are very similar to those of SN 1991T, with a slightly faster initial
decline rate after $B$ maximum in the $B$ band. The optical color curves also resemble those
of SN 1991T. The well-sampled data before the 
optical maximum show clearly that SN 1997br rises more slowly and has a wider peak
than do normal SNe Ia like SN 1994D.

An extinction estimate of $E(B-V)$ = 0.35$\pm$0.10 mag is obtained for SN 1997br by comparing the mutlicolor light
curves of SN 1997br to those of SN 1991T, and by applying an empirical relationship between
extinction and the equivalent width of interstellar Na I D lines. Lacking an
accurate distance estimate for its host galaxy, however, we cannot
obtain conclusive results for its unextinguished luminosities. 

We have compared the spectroscopic evolution of SN 1997br to that of SN 1991T and
the normal type Ia SN 1994D at various epochs.  The spectroscopic evolution of
 SN 1997br is similar to that of SN 1991T, and is peculiar 
relative to SN 1994D. Before $B$ maximum the spectra show two dominant Fe III features
centered at 4270 and 4970 \AA\, superimposed on an otherwise nearly featureless blue
continuum. As the SN ages it shows gradually the characteristic type Ia features -- Si II $\lambda$6355 and lines of other intermediate-mass elements. SN 1997br
seems to  exhibit an earlier
transition to the Fe-peak element line dominant phase than both SN 1994D and SN 1991T.
The Si $\lambda$6355 line of SN 1997br shows a shorter duration than does that of SN 1991T.
Our comparison demonstrates that the spectroscopic peculiarity of SN 1997br (and SN 1991T)
exists not only before and at optical maximum, but also in the nebular phase 
at a reduced level.

SN 1997br and other objects demonstrate the heterogeneity of SNe Ia, which
we believe is produced by
 differences in the physical conditions of the white dwarf progenitors 
and in the explosion mechanisms.
In particular, we propose a ``decelerated detonation" model which might be
promising in several ways. 

\vspace{1.5cm}
 
 We would like to thank all observational astronomers at the 2.16m telescope 
of BAO, in particular J.Y. Wei, 
X.J. Jiang, D.W. Xiu, L. Cao, J.Z. Li, X.Y. Dong and W. Liu,  for their help in our 
observations of SN 1997br. W.D. Li has benefited greatly from discussions with Prof Z.W. Li
of Beijing Normal University and Dr. Adam Riess of the University of California 
at Berkeley. We also thank Ricardo Covarrubias of CTIO for providing us with the
 calibration observations of the SN 1997br field. This work is supported by NSF grant AST-9417213 to A.V.F. and by the National Science Foundation of China.
 We are grateful to Sun Microsystems Inc. (Academic Equipment Grant Program), 
Photometrics Ltd., the National Science Foundation, and the Sylvia and 
James Katzman Foundation for donations that made KAIT possible.

\newpage

\begin{figure}
\caption{$R$-band KAIT image of the field of SN 1997br, with 
the 5 local standard stars marked. The size of the field is 5.38' $\times$
5.38'. North is up and east is to the left.}
\label{1}
\end{figure}

\begin{figure}
\caption{The $B, V, R$, and $I$ light curves of SN 1997br, along with the
 best $\chi^2$-minimizing fits to the light curves of SN 1991T.}
\label{2}
\end{figure}

\begin{figure}
\caption{The $B$ light curve of SN 1997br, together with those of SN 1991T
(Hamuy et al. 1996b), SN 1989B (Wells et al. 1994), SN 1994D (Richmond et al. 1995), 
and SN 1991bg (Leibundgut et al. 1993). All light curves are shifted in time
and magnitude to match that of SN 1997br (see text for more details).}
\label{3}
\end{figure}

\begin{figure}
\caption{Same as Fig. 3 but for the $V$ light curve.}
\label{4}
\end{figure}

\begin{figure}
\caption{Same as Fig. 3 but for the $R$ light curve. The $R$ light curve of 
SN 1991T comes from Ford et al. (1993). The $R$ and $I$ light curves of SN 
1991bg come from Filippenko et al. (1992b). }
\label{5}
\end{figure}

\begin{figure}
\caption{Same as Fig. 3 but for the $I$ light curve.}
\label{6}
\end{figure}

\begin{figure}
\caption{The $B-V$ color curve of SN 1997br, together with those of SN 1991T
(Phillips et al. 1992), SN 1989B (Wells et al. 1994), SN 1994D (Richmond et al. 1995),
and SN 1991bg (Filippenko et al. 1992b; Leibundgut et al. 1993).}
\label{7}
\end{figure}

\begin{figure}
\caption{Same as Fig. 7 but for the $V-R$ color curve.}
\label{8}
\end{figure}

\begin{figure}
\caption{Same as Fig. 7 but for the $V-I$ color curve.}
\label{9}
\end{figure}

\begin{figure}
\caption{Spectra of SN 1997br; see Table 4 for details. The phases marked are relative to the 
date of $B$ maximum. For presentation the spectra have been shifted by arbitrary amounts. }
\label{10}
\end{figure}

\begin{figure}
\caption{The day -9 spectrum of SN 1997br, together with the day -9 spectra 
of SN 1991T (Filippenko et al. 1992a) and SN 1994D (Patat et al. 1996). The spectra have been
shifted by arbitrary amounts for clarity. Line identifications
are taken from Kirshner et al. (1993), Jeffery et al. (1992), and Mazzali et al. (1995).}
\label{11}
\end{figure}

\begin{figure}
\caption{Same as Fig. 11 but for the day -4 spectrum of SN 1997br, the day
-3 spectrum of SN 1991T, and the day -3 spectrum of SN 1994D.}
\label{12}
\end{figure}

\begin{figure}
\caption{Same as Fig. 11 but for the day +10 spectrum of SN 1997br, the day
+11 spectrum of SN 1991T, and the day +10 spectrum of SN 1994D.}
\label{13}
\end{figure}

\begin{figure}
\caption{Same as Fig. 11 but for the day +14 spectra of SNe 1997br, 1991T, and 1994D.}
\label{14}
\end{figure}

\begin{figure}
\caption{Same as Fig. 11 but for the day +21 spectrum of SN 1997br, the day
+20 spectrum of SN 1991T, and the day +21 spectrum of SN 1994D.}
\label{15}
\end{figure}

\begin{figure}
\caption{Same as Fig. 11 but for the day +51 spectrum of SN 1997br, the day
+51 spectrum of SN 1991T, and the day +50 spectrum of SN 1994D.}
\label{16}
\end{figure}

\begin{figure}
\caption{Evolution of the expansion velocity as deduced from the minima of 
the Fe III $\lambda$4404, Fe III $\lambda$5129, Fe II $\lambda$4555, 
and Si II $\lambda$6355 absorptions for SN 1997br (open circles), SN 1991T
(solid circles; measured from spectra in Filippenko et al. 1992a), SN 1981B
(Branch et al. 1983), SN 1986G (Phillips et al. 1987), SN 1989B (Wells et al. 1994),
SN 1991bg (Leibundgut et al. 1993), and SN 1994D (Patat et al. 1996).}
\label{17}
\end{figure}

\end{document}



\newcommand{\bq}{\begin{equation}}
\newcommand{\eq}{\end{equation}}

\begin{deluxetable}{ccccccccccc}
\tablecaption{Magnitudes of local standard stars}
\tablehead{
\colhead{Star} & \colhead{$U$} & \colhead{$\sigma_U$} &
\colhead{$B$} & \colhead{$\sigma_B$} & \colhead{$V$} &
\colhead{$\sigma_V$} & \colhead{$R$} & \colhead{$\sigma_R$} & \colhead{$I$}
& \colhead{$\sigma_I$}  
}
\startdata
     A   & 14.646 & 0.017 &14.344 &0.004 &13.524 &0.002 &13.029 &0.004 &12.527 &0.005 \\
     B   & 18.198 & 0.204 &17.219 &0.025 &15.961 &0.008 &15.302 &0.013 &14.669 &0.016 \\
     C   & 16.873 & 0.085 &16.497 &0.017 &15.645 &0.007 &15.155 &0.011 &14.684 &0.014 \\
     D   & 17.250 & 0.090 &16.446 &0.017 &15.509 &0.007 &14.973 &0.011 &14.479 &0.013 \\ 
     E   &  $-$     &   $-$  &20.307 &0.281 &18.336 &0.046 &17.200 &0.067 &15.779 &0.071 \\
\enddata
\end{deluxetable}

\newpage

\begin{deluxetable}{cccccccll}
\tablecaption{Photometric observations of SN 1997br. Phases are relative to 
the epoch of $B$ maximum (JD$=$2450559.3)}
\tablehead{
\colhead{Date} & \colhead{J.D. - } & \colhead{Phase} &
\colhead{$B$} & \colhead{$V$} & \colhead{$R$} & \colhead{$I$} &
Telescope & Comments  \\
 DD/M/YY & 2450000.0 & (days) &&&&&&
}
\startdata
11/4/97 & 550.5 & $-9$ & 14.69 & 14.38 & 14.32 & $-$ & BAO & {} \\
12/4/97 & 551.5 & $-8$ & 14.53 & 14.24 & 14.21 & $-$ & BAO & {} \\
13/4/97 & 552.5 & $-7$ & 14.52 & 14.11 & 14.08 & $-$ & BAO & {}  \\
14/4/97 & 553.5 & $-6$ & 14.38 & 14.08 & 14.01 & $-$ & BAO & {} \\
15/4/97 & 554.5 & $-5$ & 14.25 & 14.00 & 13.92 & $-$ & BAO & poor seeing \\
16/4/97 & 555.5 & $-4$ & 14.18 & 13.96 & 13.90 & $-$ & BAO & {} \\
17/4/97 & 556.5 & $-3$ & $-$   & 13.91 & 13.87 & $-$ & BAO & {} \\
19/4/97 & 558.5 & $-1$ & $-$   & 13.86 & 13.78 & $-$ & BAO & {} \\
25/4/97 & 564.5 & $+5$ & 14.23 & 13.80 & 13.74 & $-$ & BAO & bright moon \\
26/4/97 & 564.9 & $+6$ & 14.26 & 13.77 & 13.85 & 13.70 & KAIT & \\
28/4/97 & 567.4 & $+8$ & 14.45 & 13.81 & 13.82 & $-$ & BAO & poor seeing \\
29/4/97 & 568.5 & $+9$ & 14.46 & 13.82 & 13.86 & $-$ & BAO & \\
30/4/97 & 568.8 & $+10$ & 14.57 & 13.83 & 13.94 & 13.78 & KAIT & poor seeing \\
30/4/97 & 569.4 & $+10$ & $-$  & 13.92 & 13.90 & $-$ & BAO & clouds \\
02/5/97 & 570.8 & $+12$ & 14.77 & 13.95 & 14.05 & 13.84 & KAIT & \\
02/5/97 & 571.4 & $+12$ & $-$   & 13.97 & 14.05 & $-$ & BAO & \\
03/5/97 & 572.4 & $+13$ & 15.01 & 13.99 & 14.00 & $-$ & BAO & \\
04/5/97 & 573.5 & $+14$ & 15.02 & 14.07 & 14.05 & $-$ & BAO & \\
05/5/97 & 573.8 & $+15$ & 15.07 & 14.15 & 14.14 & 13.87 & KAIT & \\
06/5/97 & 574.9 & $+16$ & $-$ & 14.21 & $-$ & $-$ & KAIT & clouds \\
07/5/97 & 575.8 & $+17$ & 15.33 & 14.24 & 14.16 & 13.85 & KAIT & \\
07/5/97 & 576.5 & $+17$ & $-$ & 14.28 & 14.09 & $-$ & BAO & \\
08/5/97 & 577.5 & $+18$ & $-$ & 14.32 & 14.14 & $-$ & BAO & clouds \\
10/5/97 & 578.8 & $+20$ & 15.67 & 14.42 & 14.20 & 13.84 & KAIT & \\
11/5/97 & 579.8 & $+21$ & 15.74 & 14.47 & 14.22 & 13.84 & KAIT & \\
13/5/97 & 581.8 & $+23$ & 15.94 & 14.57 & 14.27 & 13.86 & KAIT & \\
14/5/97 & 583.4 & $+24$ & $-$ & 14.60 & 14.30 & $-$ & BAO & clouds \\
16/5/97 & 585.4 & $+26$ & 16.13 & 14.83 & 14.35 & $-$ & BAO & \\
17/5/97 & 585.8 & $+27$ & 16.20 & 14.77 & 14.38 & 13.96 & KAIT & \\
18/5/97 & 586.8 & $+28$ & 16.24 & 14.81 & 14.40 & 13.96 & KAIT & \\
24/5/97 & 593.4 & $+34$ & 16.56 & 15.06 & 14.65 & $-$ & BAO & \\
25/5/97 & 593.8 & $+35$ & $-$ & 15.08 & 14.62 & 14.14 & KAIT & poor seeing \\
28/5/97 & 596.8 & $+38$ & $-$ & 15.25 & 14.78 & 14.28 & KAIT &  \\
28/5/97 & 597.4 & $+38$ & 16.66 & 15.20 & 14.81 & $-$ & BAO & \\
29/5/97 & 597.7 & $+38$ & $-$ & 15.28 & 14.84 & 14.32 & KAIT & \\
29/5/97 & 598.4 & $+39$ & 16.74 & 15.29 & 14.89 & $-$ & BAO & \\
31/5/97 & 599.8 & $+41$ & 16.80 & 15.36 & 14.96 & 14.45 & KAIT & \\
01/6/97 & 601.4 & $+42$ & $-$ & 15.40 & 15.05 & $-$ & BAO & \\
02/6/97 & 601.8 & $+43$ & 16.89 & 15.51 & 15.08 & 14.58 & KAIT & \\
03/6/97 & 603.4 & $+44$ & $-$ & 15.52 & 15.11 & $-$ & BAO & clouds\\
06/6/97 & 605.8 & $+47$ & $-$ & 15.62 & 15.26 & 14.77 & KAIT & \\
08/6/97 & 607.7 & $+48$ & 17.01 & 15.70 & 15.37 & 14.90 & KAIT & \\
09/6/97 & 609.4 & $+50$ & 16.96 & 15.73 & 15.45 & $-$ & BAO & \\
11/6/97 & 610.7 & $+51$ & 17.00 & 15.78 & 15.48 & 15.06 & KAIT & \\
11/6/97 & 611.4 & $+52$ & $-$ & 15.75 & 15.52 & $-$ & BAO & clouds \\
14/6/97 & 614.4 & $+55$ & $-$ & 15.80 & 15.59 & $-$ & BAO & \\
24/6/97 & 623.7 & $+64$ & $-$ & 16.05 & 15.94 & 15.60 & KAIT & \\
25/6/97 & 625.4 & $+66$ & 17.25 & 16.18 & 15.95 & $-$ & BAO & \\ 
26/6/97 & 626.4 & $+67$ & $-$ & 16.22 & 15.98 & $-$ & BAO & \\
28/6/97 & 626.7 & $+68$ & 17.21 & $-$ & $-$ & $-$ & KAIT & \\
\enddata
\end{deluxetable}

\newpage

\begin{deluxetable}{ccccc}
\tablecaption{Peak apparent magnitudes of SN 1997br}
\tablehead{
\colhead{Filter} & \colhead{$B$} & \colhead{$V$} &
\colhead{$R$} & \colhead{$I$} 
}
\startdata

Epoch of Max. & Apr 20.3$\pm$1.5 & Apr 23.4$\pm$1.0  &  Apr 23.2$\pm$1.0  & Apr 19.8$\pm$2.0 \\
Julian Date of Max. & 559.1$\pm$1.5& 562.2$\pm$1.0&562.0$\pm$1.0&558.6$\pm$2.0 \\
 ( - 2450000.0) &&&& \\
Magnitude at Max. & 14.06$\pm$0.10 &13.74$\pm$0.06 &  13.64$\pm$0.10 & 13.50$\pm$0.20
\enddata
\end{deluxetable}
{\vspace*{18.0cm}

\newpage

\begin{deluxetable}{ccccl}
\tablecaption{Journal of Spectroscopic Observations}
\tablehead{
\colhead{UT Date} & \colhead{J.D. $-$ } & \colhead{Phase} &
\colhead{range} & \colhead{equipment} \\
 DD/M/YY &  $2450000.0$   & (days) & \AA\ & 
}
\startdata
11/4/97 & 550.2 & $-9$ & 3800$-$8000 & BAO$-$2.16 \\
12/4/97 & 551.2 & $-8$ & 4000$-$8000 & BAO$-$2.16 \\
13/4/97 & 552.2 & $-7$ & 3800$-$8200 & BAO$-$2.16 \\
14/4/97 & 553.2 & $-6$ & 4000$-$8400 & BAO$-$2.16 \\
16/4/97 & 554.8 & $-4$ & 3200$-$10200 & Lick$-$3.0 \\
16/4/97 & 555.2 & $-4$ & 3600$-$8200 & BAO$-$2.16 \\
28/4/97 & 567.1 & $+8$ & 3400$-$9400 & BAO$-$2.16 \\
30/4/97 & 569.2 & $+10$ & 3400$-$10600 & BAO$-$2.16 \\
02/5/97 & 571.1 & $+12$ & 4000$-$8800 & BAO$-$2.16 \\
04/5/97 & 573.2 & $+14$ & 3800$-$8800 & BAO$-$2.16 \\
08/5/97 & 577.1 & $+18$ & 4000$-$10200 & BAO$-$2.16 \\
11/5/97 & 509.8 & $+21$ & 3000$-$7000 & Lick$-$3.0 \\
14/5/97 & 583.1 & $+24$ & 4000$-$8000 & BAO$-$2.16 \\
28/5/97 & 597.1 & $+38$ & 3800$-$8400 & BAO$-$2.16 \\
06/6/97 & 605.8 & $+47$ & 3400$-$10400 & Lick$-$3.0 \\
10/6/97 & 610.1 & $+51$ & 3800$-$8400 & BAO$-$2.16 \\
26/6/97 & 626.1 & $+67$ & 3800$-$8200 & BAO$-$2.16 \\
10/7/97 & 639.7 & $+81$ & 4000$-$10000 & Lick$-$3.0
\enddata
\end{deluxetable}